\documentstyle[12pt,epsfig,rotating]{article} 
\newcommand{\ptm}{p_{t,\mbox{\footnotesize max}}}

\newlength{\dinwidth}
\newlength{\dinmargin}
\begin{document}
\setlength{\unitlength}{1mm}
\begin{titlepage}
\begin{flushleft}
%
%
{\tt DESY 96-224    \hfill    ISSN 0418-nnnn} \\
{\tt October 1996}                  \\
       
\end{flushleft}
\vspace*{4.cm}
\begin{center}
\begin{Large}
 
{\boldmath \bf Scale Influence on the Energy Dependence
               of Photon-Proton Cross Sections } \\
 
\vspace{1.cm}
{H1 Collaboration}    \\
\end{Large} \end{center}
\vspace*{2.5cm}
\begin{abstract}
\noindent
 The scale dependence of the evolution
  of photoproduction cross sections with the photon-proton 
centre of mass  energy $W$ 
  is studied 
using  low $Q^2 < 0.01$ GeV$^2$  $e^+p$ interactions 
collected by the H1 experiment at HERA.
The value of the largest transverse momentum of a 
charged particle 
 in the photon fragmentation region is used to define  the 
hard scale.
The slope of the $W$ dependence of the cross section is observed to increase 
steeply with increasing transverse momentum. The result is
compared  to   measurements
 of the $Q^2$ evolution  of the $W$ dependence of the
virtual photon-proton cross section.
 Interpretations in terms of QCD and in terms of Regge phenomenology
are discussed.
\end{abstract}
%
%
\end{titlepage}
 S.~Aid$^{13}$,                   
 M.~Anderson$^{23}$,              
 V.~Andreev$^{26}$,               
 B.~Andrieu$^{29}$,               
 A.~Babaev$^{25}$,                
 J.~B\"ahr$^{36}$,                
 J.~B\'an$^{18}$,                 
 Y.~Ban$^{28}$,                   
 P.~Baranov$^{26}$,               
 E.~Barrelet$^{30}$
% ...

% ...
$ ^2$ III. Physikalisches Institut der RWTH, Aachen, Germany$^ a$ \\
% ...
$ ^{13}$ II. Institut f\"ur Experimentalphysik, Universit\"at Hamburg,
                                     Hamburg, Germany$^ a$  \\
% ...
$ ^{18}$ Institute of Experimental Physics, Slovak Academy of
                Sciences, Ko\v{s}ice, Slovak Republic$^{f, j}$\\
% ...
$ ^{23}$ Physics Department, University of Manchester,
                                          Manchester, UK$^ b$\\
% ...
$ ^{25}$ Institute for Theoretical and Experimental Physics,
                                                 Moscow, Russia \\
 $ ^{26}$ Lebedev Physical Institute, Moscow, Russia$^ f$ \\
% ...
$ ^{28}$ LAL, Universit\'{e} de Paris-Sud, IN2P3-CNRS,
                            Orsay, France\\
 $ ^{29}$ LPNHE, Ecole Polytechnique, IN2P3-CNRS,
                             Palaiseau, France \\
 $ ^{30}$ LPNHE, Universit\'{e}s Paris VI and VII, IN2P3-CNRS,
                              Paris, France \\
% ...
$ ^{36}$ DESY, Institut f\"ur Hochenergiephysik,
                              Zeuthen, Germany$^ a$\\
% ...
$ ^a$ Supported by the Bundesministerium f\"ur Bildung, Wissenschaft,
        Forschung und Technologie, FRG,
        under contract numbers 6AC17P, 6AC47P, 6DO57I, 6HH17P, 6HH27I,
        6HD17I, 6HD27I, 6KI17P, 6MP17I, and 6WT87P \\
 $ ^b$ Supported by the UK Particle Physics and Astronomy Research
       Council, and formerly by the UK Science and Engineering Research
       Council \\
% ...
$ ^f$ Supported by the Deutsche Forschungsgemeinschaft \\
% ...
$ ^j$ Supported by VEGA SR grant no. 2/1325/96 \\
 
%
\newpage
\section{Introduction}
\noindent
At high energies
measurements of the energy dependence of real photon-proton cross sections
are consistently described by the power law 
$\sigma^{tot}_{\gamma p} \propto (W^2)^\lambda$, where $W$ is the 
$\gamma p$ centre of mass energy.
The power is
$\lambda \approx 
0.08$ \cite{stot_H1,stot_Z}
 and is found to be universal for all photon or hadron 
 interactions with protons \cite{DL}.
The virtual photon-proton cross section however, which is 
at large $W$ related to the
structure function $F_2$ via $\sigma^{tot}_{\gamma^*p}\approx
 4\pi^2\alpha/Q^2 \cdot F_2(W^2,Q^2)$, is found
to rise fast with increasing $W^2$
for $Q^2$ values larger than a few 
GeV$^2$ \cite{F2H,F2Z}. Here $Q^2$ is the virtuality of 
the photon and $\alpha$ the 
fine structure  constant. At large $W$ the relation $W^2 = 
Q^2/x$ holds, with $x$ the deep inelastic 
scattering (DIS) variable Bjorken-$x$.
In the HERA regime fits of the form 
$  \sigma^{tot}_{\gamma^*p} ~\propto ~(W^2)^\lambda$
lead to values of $\lambda$  which rise from 0.2 to 0.4 
 in the 
$Q^2$ range from $1.5$ to $10^3$ GeV$^2$~\cite{F2H}.

In Regge theory the energy dependence of the 
total cross sections is specified by the intercept,
which is simply $1+\lambda$, of the pomeron, the leading
exchanged trajectory.
 It is  argued
(see e.g. \cite{bartels,levy}) that
the change of slope parameter $\lambda$ with increasing $Q^2$ is associated 
with a transition from a non-perturbative `soft' to a perturbative `hard' 
regime. 
Indeed, in the domain where  $Q^2$ is larger than a few GeV$^2$
 and perturbative QCD 
 can be safely applied,  it predicts an increase of  $\lambda$ with $Q^2$. 
In the domain of `soft' interaction phenomenology  the change of the slope 
parameter can be interpreted
in  Reggeon Field Theory 
  as a reduction of screening corrections with increasing
$Q^2$~\cite{CKMT}.
Presently   Reggeon Field Theory and perturbative QCD  are  
 two complementary approaches which successfully 
 describe  physics processes in 
different regimes, but  the 
transition region between these soft and hard 
regimes is  poorly understood.
HERA provides an ideal testing ground to study the transition from 
soft to hard physics by comparing real and virtual photon-proton
collisions in the same  experiment.
Comparisons of the energy dependence  of the cross section or of the 
final states properties (e.g. multiplicities, energy flows, inclusive particle 
distributions)
allow   more insight to be gained into the dynamics of strong interactions
and can be helpful in the development of a common underlying picture
in terms of an effective field theory (see e.g. \cite{Lipatov}).

\begin{figure}[htb]
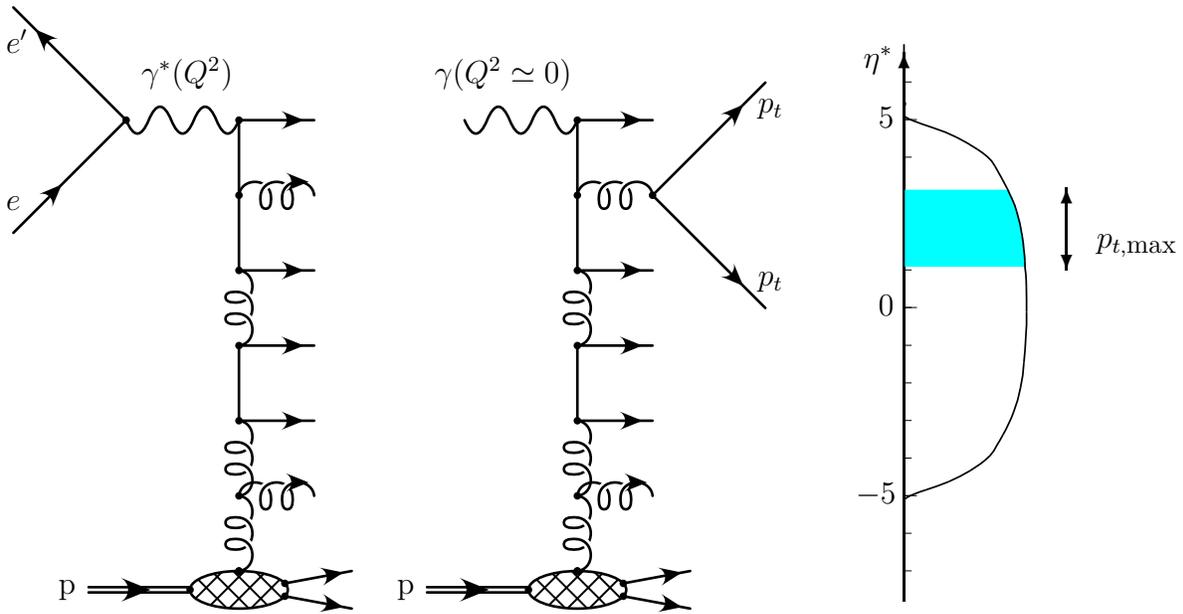
 \centering 
\begin{picture}(100,90)(05,-05)
\thicklines       
\put(17,70){$\gamma^* (Q^2)$} \put(56,70){$\gamma (Q^2 \simeq 0)$}
\put(-1,53){$e$}  \put(-1,74){$e'$}  \put(6,2){p}  \put(51,2){p}
\put(99,43){$p_t$}  \put(99,66){$p_t$}
\put(115,64){5} \put(115,39){0} \put(112,14){$-$5} \put(113,72){$\eta^*$}
\put(118.4,1){\vector(0,1){73}} \put(144,48){$\ptm$}
\put(140,49){\vector(0,1){7}}  \put(140,49){\vector(0,-1){4}}
\epsfig{file=Fig0_FD.ps,width=100mm}
\end{picture}
\begin{picture}(40,90)(-12,-10)
\epsfig{file=Fig0_sketch1.eps,height=70mm}
\end{picture}
\caption{Illustration of diagrams for 
         a DIS event and photoproduction event with 
         high $p_t$ particles in the final state. The sketch on the
         right side illustrates an inclusive charged particle
         pseudorapidity distribution in the $\gamma p$ centre of mass system,
         with the range used to search for high $p_t$ tracks.
         For real photon interactions all contributing 
    processes (direct, resolved, vector meson production, ...) are considered.}
\label{FD}
\end{figure}

In a previous publication  energy flow in the photon fragmentation
region was studied~\cite{flow}.
In this paper 
we investigate  the energy behaviour of the 
cross    section for photoproduction  events containing charged 
particles with high transverse momentum, $p_t$, and compare it with the
energy behaviour of the total inclusive $\gamma^*p$ cross section.
The presence of a high $p_t$ charged particle 
allows a scale to be identified which  can be varied in a similar way as the 
scale in
virtual photon interactions, and the effect on the energy dependence of the 
cross section  can be studied.
A continuous coverage of  both
the soft and hard scattering domains becomes possible.
To emulate the kinematical configuration of
a $\gamma^* p$ collision  as closely  as possible the  high $p_t$ charged 
track is required to be in the
photon fragmentation
region, which corresponds 
at the same time  to 
the largest acceptance  region of the H1 tracking detectors.
The idea of the analysis is illustrated in Fig.\ref{FD}, where both the 
virtual and real photon processes are shown, as well as the 
region of track acceptance.
%
%

\section{Detector and Data Sample}
The data for this analysis were collected by the H1 experiment
in 1994,
when HERA operated with  27.5 GeV positrons and 820 GeV protons, thus
producing $e^+p$ interactions at a centre of mass energy of $\sqrt{s}=300$ GeV.
A detailed description of the H1 apparatus can be found elsewhere~\cite{H1NIM}.
The detector covers almost 4$\pi$ of the solid angle and consists, 
moving outwards from the interaction point, 
of tracking detectors, calorimeters and muon detectors. Along the
beam line several other dedicated detectors, such as the luminosity system,
have been installed.
This analysis  makes use of
the luminosity system and the central tracking detector.

The scattered positron was detected in the electron tagger of the 
H1 luminosity system.
The system consists of two TlCl/TlBr crystal calorimeters, installed
in the HERA tunnel, each having a resolution of 
$\sigma(E)/E=0.1/\sqrt{E}$ with $E$ in GeV. 
The electron tagger is located at $z=-33$ m and the photon tagger 
at $z=-103$ m from the interaction point in the direction of outgoing
positron beam.
The  energy scale of both calorimeters is known to better than 1.5\%
\cite{QED}. The acceptance of the electron tagger is confined 
to the kinematical range of $0.2 < y < 0.8$ and
$Q^2 < 0.01$ GeV$^2$. Here $y$ is the fractional energy of the photon and
can be well approximated by $y=1-E'_{e}/E_e$, with 
 $E_e$,  $E'_{e}$  the energies
of the initial and scattered positrons respectively.
The photon tagger 
         accepts  photons with 
$\pi - \theta < 0.45$ 
mrad. Here $\theta$ is defined with respect
to the incident proton direction.
 
The charge and momentum of charged particles  
were measured by two coaxial cylindrical drift chambers
(central jet chamber, CJC), which cover the polar angular range
$ 15^\circ < \theta < 165^\circ$. 
A superconducting solenoid provides a uniform magnetic field of 1.15 T
parallel to the beam axis in the tracking region.
The  $p_t$ and $\theta$ resolutions of the CJC are
$\sigma_{p_t}/p_t \approx 0.009 \cdot p_t$ [GeV] $\oplus$ 0.015 and
$\sigma_{\theta} = 20$ mrad respectively. 

Photoproduction events 
were triggered by a coincidence of  an  energy deposit larger
than  $\sim 5$ GeV in the electron tagger
and one or more track candidates with $p_t > 0.4$ GeV 
from the CJC trigger \cite{cjctrig}.
In order to reduce the non-$ep$ background contamination and to 
ensure  good 
reconstruction of the event kinematics, the following selection criteria
have been applied for this analysis:
\begin{itemize}
  \item  The event vertex reconstructed from the tracks of charged particles
         was required to lie within $\pm 30$ cm of the mean $z$-position
         of the interaction point, corresponding to a $3\sigma$ cut
         on the interaction region defined by the longitudinal
         size of the proton bunch.
  \item  In order to suppress random coincidences between the
         high rate of Bethe-Heitler events
         $ep \rightarrow ep\gamma$ and
         $p$-gas background in the main H1 detector,
         events were rejected if a photon with an energy 
         $E_{\gamma} > 2$ GeV
         was detected in the photon tagger.
         This cut also reduces  QED radiative corrections to the
         $ep$ Born cross section~\cite{stot_H1}.
  \item  The $y$ range was limited to 
         $0.25 < y < 0.70$
         to avoid regions of low electron tagger acceptance~\cite{stot_H1}. 
         This corresponds to  photon-proton collision
         energies of $150 < W = \sqrt{ys} < 250$ GeV.  
  \item  To ensure a good measurement of the scattered positron energy,
         a fiducial cut   was applied
         to reject events with a positron detected 
         close to the acceptance  boundaries  of the electron tagger.
\end{itemize}
In total  $1.8 \cdot 10^6$ events satisfy the above selection criteria,
from a data sample which corresponds 
to an integrated luminosity of $\cal{L}$ $ = 2.6$ pb$^{-1}$.
The residual background contamination was estimated to be $(2.4 \pm 0.5)\%$.

The  data  were divided into several sub-samples  according
 to the charged particle with the maximum transverse momentum, $\ptm$, 
 found
 in the pseudorapidity range $1.1 <  \eta^* < 3.1$.
Here  pseudorapidity is defined in the photon-proton 
centre of mass system as $\eta^* = -\ln\tan(\theta^*/2)$
with the polar angle $\theta^*$ measured with respect to the photon direction. 
The  pseudorapidity interval chosen corresponds
 to a region in the detector which guarantees a  good
track measurement in the CJC and
a uniform track acceptance over  the whole $W$ range studied.
Only those tracks were considered
which have $p_t > 0.15$ GeV, 
 are fitted to the primary event vertex 
and further satisfy  a number of quality criteria detailed in 
Ref.~\cite{Tracks}.
The systematic error due  to the track selection was
estimated by varying these quality criteria in a wide range and 
repeating the complete analysis.
In total eleven bins in $\ptm$ have been defined, 
covering the $\ptm$ range from 0.5 
to 7 GeV. The bin sizes were determined to have adequate statistics
and such that the size is always larger than
four times  the $p_t$ resolution in the bin to minimize  migration 
effects.
The values of the $\ptm$ boundaries for the bins are indicated in Fig.~2.

The  track reconstruction efficiency, the contamination
from secondary vertices and  decays close to the primary vertex, 
and  residual bin-to-bin migrations have been determined from
Monte Carlo studies. Events were generated with the 
PYTHIA model~\cite{PYTHIA} and
passed through the full H1
detector simulation. All simulated events were subjected  
to the same reconstruction and selection procedure as the real data.
The migration effects were found to be small, between 10\% and 20\%,
and the ratio of the number of reconstructed  to the 
number of  generated events
to be close to one ($\pm 4\%$) in  all $\ptm$ bins.

\section{Results and Discussion}
In $ep$ collisions, for any sub-class $i$ of photoproduction events
the energy dependence $\sigma^i_{\gamma p}(W)$ can be extracted
from the corresponding differential $ep$ cross section using the
Weizs\"{a}cker-Williams formula~\cite{WWA} for the photon flux $F(y,Q^2)$.
After integrating over $Q^2$, the corresponding relation reads:
\begin{equation}
  \frac{dN^i_{ep}}{dy} = (1+\delta_{RC})~{\cal L}~F(y)~A(y)~\epsilon^i(y)
                       ~\sigma^i_{\gamma p}(W),        
\label{Sigma} 
\end{equation}
where $N^i_{ep}$ denotes the observed number of events belonging to
class $i$, $\cal L$ is the integrated luminosity,
$A(y)$ is the acceptance of the electron tagger and
$\epsilon^i(y)$ is the efficiency of the trigger and selection criteria.
The factor $(1+\delta_{RC})$ is a correction for the cross section
for  QED radiation 
to obtain the $ep$ Born cross section.

The main  systematic uncertainty,  which so far prevents an accurate
measurement of the energy dependence of the total photoproduction
cross section at HERA, results from the   large variation of the
tagging acceptance $A(y)$ (between 0.15 and 0.8 in the $W$
range of interest), and which, in addition, strongly depends on the beam optics
\cite{stot_H1,QED}. However,
the {\it relative} energy dependence
for different photoproduction event classes can be determined
 with much better precision.
Hence we consider the ratio
\begin{equation}
  R^i(y) = \frac{\sigma^i_{\gamma p}(W)}{\sigma^{tot}_{\gamma p}(W)} = 
      \frac{dN^i_{ep}/dy}{dN_{ep}/dy} \cdot
      \frac{\epsilon(y)}{\epsilon^i(y)} \equiv 
      \frac{\tilde{R}^i(y)}{r_{\epsilon}^i(y)},       
\label{Ratio} 
\end{equation}
where 
 $\epsilon$ and $\epsilon^i$ denote the efficiencies
 for all events and for the events in class $i$ respectively, and 
$\tilde{R}$ is the uncorrected ratio.             
All efficiencies in~(\ref{Ratio})
cancel  except for the trigger efficiency. Hence
$r_{\epsilon}^i$ is the ratio of the trigger efficiency for the 
sample $i$ to that  of the full sample.

Applying this method  
the  $W$
dependence of the uncorrected ratio $\tilde{R}$ for each bin $i$ of
$\ptm$ is shown in Fig.~\ref{YFIT} together with             
a power-law fit  of the form 
$\tilde{R} \propto W^{2 \cdot \tilde{\lambda}(\ptm)}$.
\setlength{\unitlength}{1mm}
%
%
\begin{figure}[h] \centering        
\begin{picture}(140,140)(0,0)
\put(105,118){$0.5<\ptm<0.7$}
\put(105,110){$0.7<\ptm<0.8$}
\put(105,100){$0.8<\ptm<1.0$}
\put(105,092){$1.0<\ptm<1.3$}
\put(105,084){$1.3<\ptm<1.6$}
\put(105,077){$1.6<\ptm<2.0$}
\put(105,069){$2.0<\ptm<2.5$}
\put(105,061){$2.5<\ptm<3.2$}
\put(105,049){$3.2<\ptm<4.0$}
\put(105,040){$4.0<\ptm<5.0$}
\put(105,032){$5.0<\ptm<7.0$}
\put(113,24){$[$GeV$\,]$}
\put(45,121){$\times~4$} \put(45,113){$\times~5$} \put(45,104){$\times~1.5$} 
\begin{sideways} \put(72,0){\LARGE \~{R}} \end{sideways}  
\put(45,7){\large $y = W^2/s$}
 \put(-10,0){\epsfig{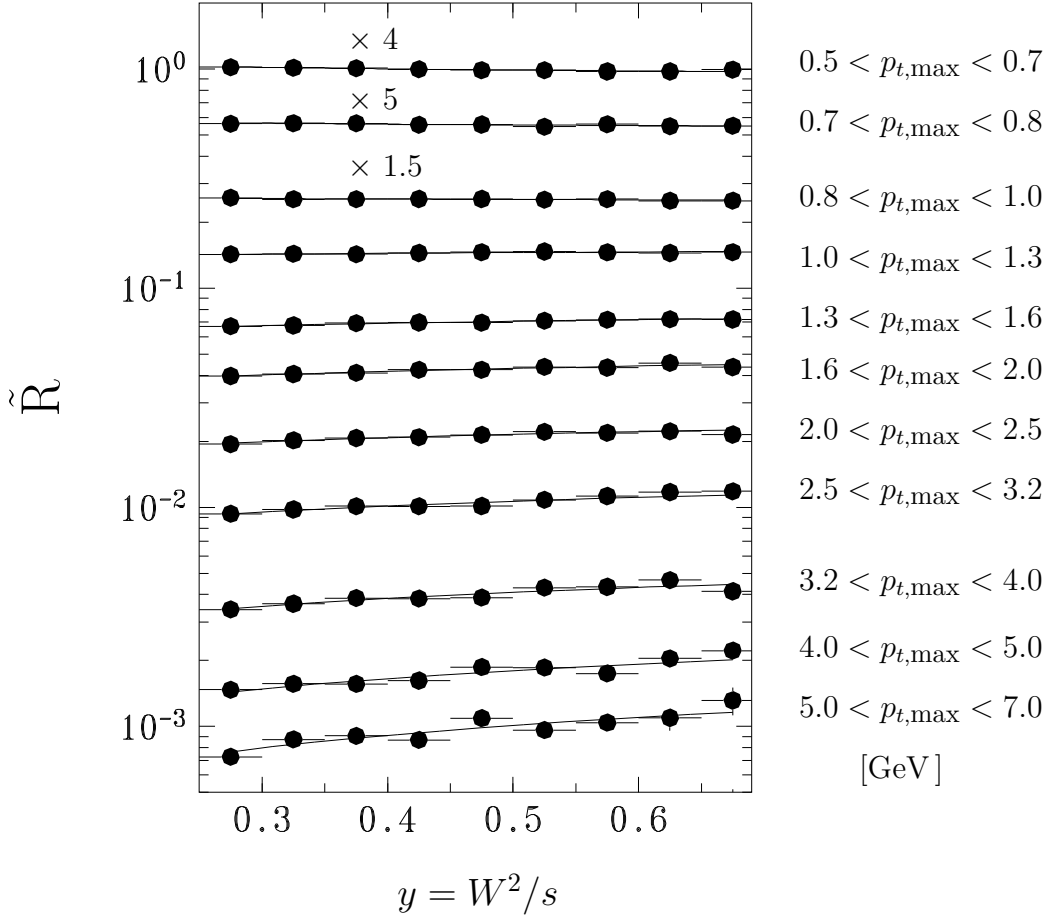}}
\end{picture}                              
\caption{The energy dependence of the ratio $\tilde{R}$
         in different bins of the
         $\ptm$ of the charged particles produced in the
         pseudorapidity window $1.1 < \eta^* < 3.1$.
         The curves represent a  fit
         of the form $\tilde{R} \propto W^{2 \cdot \tilde{\lambda}(\ptm)}$
         motivated by the Regge formalism.
         For visibility the data of the three lowest $\ptm$ bins are multiplied
         with the factors indicated in the figure.}
\label{YFIT}
\end{figure}

In order to obtain the energy dependence of the partial cross sections,
defined in a $\ptm$ bin with width $2\Delta$,
$$\sigma^i_{\gamma p}(W)\equiv
\sigma_{\gamma p}(W,\ptm^2) = \int_{\ptm-\Delta}^{\ptm+\Delta}
d\sigma_{\gamma p}(W)/d\ptm \cdot d\ptm$$ in (\ref{Ratio})
the parameters $\tilde{\lambda}(\ptm)$ have been corrected for  
trigger efficiency
$r^i_{\epsilon}(y)$ and account has been taken of 
the energy dependence of the total cross section
in the denominator of (\ref{Ratio}):
\begin{equation}
 \lambda_{\gamma p}(\ptm) = \tilde{\lambda}(\ptm)-\lambda_{\epsilon}(\ptm)+
\lambda_{I\!\!P}
 \label{Lam}
\end{equation}
where  $\lambda_{\epsilon}(\ptm)$
is the  trigger efficiency correction  and $\lambda_{I\!\!P} = 
0.08 \pm 0.03$.
 The rather conservative error assigned to 
$\lambda_{I\!\!P}$
accounts for 
the invisible part of the total $\gamma p$ cross section 
with the data selection of this analysis (mainly elastic
and low-mass diffractive dissociation processes)~\cite{stot_H1}.
The correction due to the trigger efficiency was obtained directly from the
data, using a sub-sample taken with a trigger based solely
on   calorimetric quantities.
Correlated systematic errors of the $r^i_\epsilon(y)$  largely cancel in
the measurement of the slope parameter  $\lambda_{\gamma p}$.
The contributions of the systematic errors
to the total error  $\delta\lambda_{\gamma p}$, 
excluding  the overall uncertainty in 
$\lambda_{I\!\!P}$, are summarized in  Table 1. The rightmost column 
gives  the systematic error 
resulting from the precision of the $W^2$ reconstruction and
the stability of the power-law fit to   variations 
of the $W^2$ range used.  
\begin{table}[htb]      
    \caption[]{                                                   
     Values and the 
     systematic error summary for the measurement of the slope
parameter  $\lambda_{\gamma p}(\ptm)$.
     The overall systematic error $\delta\lambda_{I\!\!P}=0.03$ is not included
     in this table, but is included in Fig.\ref{SLOPE}a.}
\begin{center}
\begin{tabular}{|r|r||r|r|r|r|}
  \hline 
 Bin & $<\ptm^2>$ & $\lambda_{\gamma p}(\ptm) \pm \delta 
\lambda_{\gamma p}(stat) $ &
  \multicolumn{3}{|c|}{Systematic errors} \\ \cline{4-6}
   & [GeV$^2$] & & track sel. & trigger & $W^2$ fit \\
   \hline
  1 &   0.36  & $0.080 \pm 0.006$ & 0.001 & 0.02 & 0.006 \\
  2 &   0.56  & $0.121 \pm 0.009$ & 0.001 & 0.01 & 0.008 \\
  3 &   0.80  & $0.157 \pm 0.007$ & 0.001 & 0.01 & 0.007 \\
  4 &   1.29  & $0.211 \pm 0.008$ & 0.002 & 0.01 & 0.004 \\
  5 &   2.06  & $0.267 \pm 0.012$ & 0.001 & 0.01 & 0.004 \\
  6 &   3.14  & $0.334 \pm 0.015$ & 0.001 & 0.01 & 0.007 \\
  7 &   4.90  & $0.344 \pm 0.022$ & 0.001 & 0.01 & 0.003 \\
  8 &   7.77  & $0.407 \pm 0.031$ & 0.001 & 0.01 & 0.006 \\
  9 &  12.46  & $0.484 \pm 0.051$ & 0.015 & 0.02 & 0.008 \\
 10 &  19.56  & $0.565 \pm 0.079$ & 0.036 & 0.03 & 0.004 \\
 11 &  33.43  & $0.603 \pm 0.105$ & 0.071 & 0.07 & 0.062 \\
\hline  
\end{tabular}
\end{center}
\end{table}

\begin{figure}[htb] \centering        
\begin{picture}(177,120)(3,-4)
\large                                                      
\begin{sideways} \put(93,-3){slope $\lambda$} \end{sideways}  
\put(37,102.5){H1 $\gamma p$ data}
\put(37,95.8){Phase space}       
\put(37,89.0){PYTHIA} 
\put(109,102.5){H1 DIS data} \put(109,096.0){GRV94} 
\put(109,89.0){MRS-D0$^{\prime}$}
\put(75,103){(a)} \put(148,103){(b)} 
\normalsize
\epsfig{file=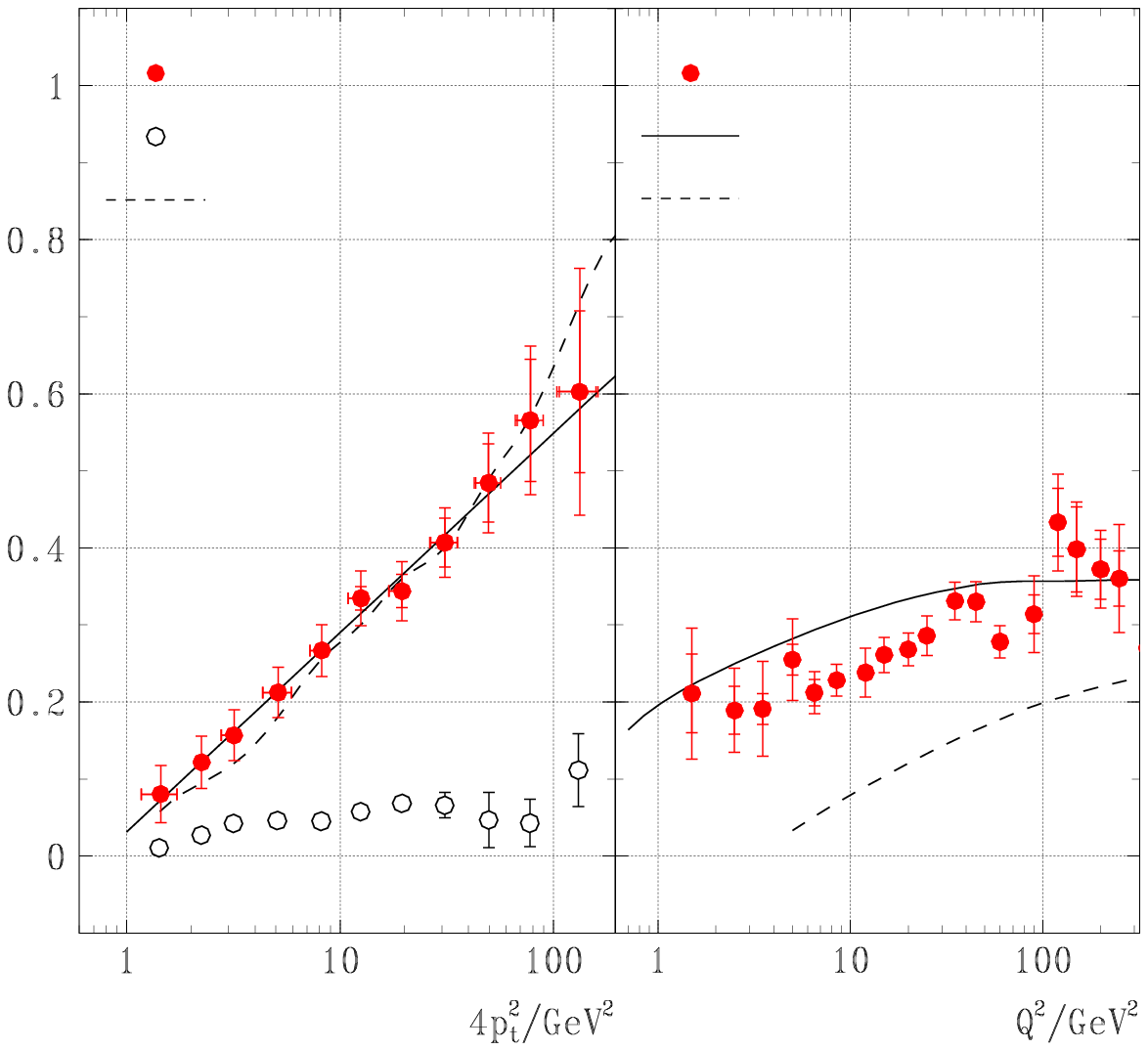,width=122mm,height=177mm,angle=90}
 \put(-108,-7){\epsfig{file=white.eps}} \put(-44,-7){\epsfig{file=white.eps}}
 \put(-117,1){\large $4\ptm^2$/GeV$^2$} \put(-40,1){\large $Q^2$/GeV$^2$}
\end{picture}
\caption{ a) The scale dependence of the slope $\lambda_{\gamma p}(\ptm)$ in 
             photoproduction (full points).
            The full line represents the result of a linear fit
            $\lambda_{\gamma p}(\ptm) = C \cdot \ln(4\ptm^2) + \lambda_0$ 
            to the data. The dashed line is the prediction from PYTHIA and 
            the open points are the prediction from a 
            longitudinal phase space model.
         b) The scale dependence of the slope $\lambda_{DIS}(Q^2)$ in 
            deep inelastic data  compared to different parametrisations
            of the proton structure function $F_2$: full line GRV, dashed
            line MRS-D0$^{\prime}$.
            For both figures inner error bars 
            show the statistical errors;  full
            error bars correspond to statistical and systematic errors
            added in quadrature.}
\label{SLOPE}
\end{figure}

The final corrected values of $\lambda_{\gamma p}(\ptm)$ are 
plotted in Fig.\ref{SLOPE}a 
against the scale which is chosen to be $(2\ptm)^2$.
The statistical errors dominate the  precision of the $\lambda_{\gamma p}
(\ptm)$ 
measurement
in the high-$p_t$ range ($\ptm > 3$ GeV).
%
%

A fit to the data of the form $\lambda_{\gamma p}(\ptm) = C \cdot \ln(4\ptm^2) + \lambda_0$
with full errors except for $\delta\lambda_{I\!\!P}$
(full straight line on Fig.\ref{SLOPE}a) gives a $\chi^2/ndf = 0.3$ with 
$C=0.112 \pm 0.007$ and $\lambda_0 = 0.031 \pm 0.014$.
Note that when the scale $4\ptm^2$ is multiplied with a constant, this
does  not change the slope $C$ of the scale dependence, but
only influences the value of $\lambda_0$.
Thus the change of $\lambda_{\gamma p}(\ptm)$ with the scale has  physical   significance.

Fig.\ref{SLOPE}a shows also   predictions of  a longitudinal
phase space model (LPS) and of the  PYTHIA model~\cite{PYTHIA}.
The PYTHIA model is  based on 
perturbative QCD for the hard scattering processes, but also includes 
a Regge inspired soft interaction component.
For the LPS model particles were generated with a uniform distribution  in 
rapidity, taking into account  energy-momentum constraints, and 
with a transverse momentum distribution 
$   d\sigma/dp^2_t \propto (p_t^2+p_0^2)^{-n}$
   with $p_0 = 0.5$ GeV and $n=2.9$, tuned to describe the charged particle
   multiplicity and $p_t$ spectra of the data in the 
   photon fragmentation region.
   As expected,    phase space effects
start to become important only at very large $\ptm$ values
and  cannot explain the  trend  observed in the data.
PYTHIA, on the other hand, describes the data
well. In studying the different ingredients in this model it turns out that
the main part of the rise can be explained
 by the integration of the 
 leading order  matrix elements over the available phase space.
This is in agreement with expectations reported in \cite{halzen}.
Since this integration is regulated by the quantity $p_t^2/\hat{s}$, 
with $\hat{s}$ the invariant mass 
squared of the hard scattering subsystem, the rise 
 for a given $\ptm$ value will depend on the $W$ range studied.
Parton showers, simulating next to leading order effects,
as well as multiple parton-parton scattering, account for
at most  20\% of the measured $\lambda_{\gamma p}(\ptm)$. 
The effects of using different  proton and photon structure
functions  are small.
Hence the change of $\lambda_{\gamma p}(\ptm)$
 is not associated with the growth of the gluon density
in the proton at small-$x$, contrary to naive expectation.

In Fig.\ref{SLOPE}b the measurement  of $\lambda$ in 
 $  \sigma^{tot}_{\gamma^*p} ~\propto ~(W^2)^{\lambda_{DIS}(Q^2)}$ is 
shown for the virtual-photon proton cross section as function of the 
virtuality $Q^2$~\cite{F2H}.
For DIS the $Q^2$ dependence of 
$\lambda_{DIS}(Q^2)$ is qualitatively   described
by perturbative QCD, as shown in Fig.~3b  by the  calculations 
for two  quite different
proton structure function parametrisations~\cite{GRV,MRSD0}.
Both  are based on the 
DGLAP evolution equations~\cite{DGLAP} but assume a different 
$x$-behaviour at the starting scale, $Q_0^2$, of the evolution.
   At $Q^2 \sim 4 $ GeV$^2$ the MRS-D0$^{\prime}$
 parton densities are constant in $x$
 $(\lambda \sim 0)$, 
   while for  GRV they are  singular at $x = 0$ $ (\lambda > 0)$. This 
difference
   is reflected in the predictions
    for the GRV and MRS-D0$^{\prime}$ parametrisations  in  Fig.~3b. 
However, 
  both predict a significant 
   rise of $\lambda_{DIS}(Q^2)$ with $Q^2$, due to  the singular behaviour
    of the box diagram $g\rightarrow q\overline{q}$
of the DGLAP evolution kernel~\cite{Collins}.
In detail neither of the two calculations agrees very well 
with the data which are
in between the predictions.
Note that the MRS-D0$^{\prime}$
 distributions were found to be in disagreement with 
the structure function data itself~\cite{F2H,F2Z}.

A  qualitative similarity is seen between the $W$ dependence
of the cross sections for   
  high-$p_t$
   particle photoproduction and for deep inelastic scattering:
they both show a significant rise of $\lambda_{\gamma p}(\ptm)$ and 
$\lambda_{DIS}(Q^2)$ with increasing scale. 
 The use of different
scales 
in the study of this effect in the two processes prevents 
however a direct
quantitative 
comparison. The range of the scale is varied from 1 GeV$^2$ 
to 100 GeV$^2$ and
within that  range the rise of the slope parameter is  stronger for
$\sigma_{\gamma p}(W,\ptm^2)$.
In view of the  discussion above the rise 
in both processes
 can be understood as follows:

\begin{itemize}
\item
      Perturbative QCD : as argued above, the rise of the cross section
      for photoproduction events 
      with a high $p_t$ particle  can be  explained in terms
      of parton-parton scattering. On the other hand, the 
      rise of the slope parameter due to  QCD evolution in DIS  can be traced
      to the  singular behaviour of 
      the box diagram describing the elastic $\gamma^*p$
      scattering amplitude. This diagram gives the contribution to the 
      $\gamma^*p$ cross section of 
      photon-parton scattering with  gluon or quark radiation
      and can be expected to have  similar properties 
      as the diagrams
      for parton-parton scattering in high $p_t$ photoproduction.
    It can be shown~\cite{Field}
    that  the photon-gluon fusion diagram
    leads to a  rise of $F_2$ which is  compatible with the 
    HERA data~\cite{Buchmuller}. Hence, in  view of Fig.~1, the
    cross section evolution with the scale ($p_t^2$ in photoproduction and
    $Q^2$ in DIS) can then be readily explained by  leading
    order partonic scattering
    contributions.
\item
     Regge approach: in  Reggeon Field Theory, 
     high energy particle interactions are  described by pomeron exchange 
and 
  $\lambda$ is related to the pomeron intercept. 
     At high energies  the unitarity corrections to the 
     one-pomeron exchange, which 
    effectively reduce the cross section
     value, become important.
     It has been  shown~\cite{Kada} that these reduce the value of $\lambda$.
     In both cases studied here 
     the  cross sections are far below the unitarity 
     bound~\cite{levin_fr}
 and in a region  where unitarity corrections become gradually
     smaller~\cite{CKMT}. Hence the cross section can rise
     faster with $W$ for either increasing $p_t^2$ or $Q^2$.
     The data shows that this is indeed the case and the development of the 
rise is similar, albeit not identical, for both cases.
\end{itemize}

In short, an
 interesting similarity is observed
in the evolution of the energy dependence
of $\sigma_{\gamma p}(W,\ptm^2)$ and $\sigma_{\gamma^*p}^{tot}(W,Q^2)$, 
in qualitative agreement with leading order perturbative 
QCD and Regge  Field Theory expectations.
 When decreasing the scale 
from large values towards 1 GeV$^2$, which 
corresponds to a $\ptm$ of 0.5 GeV, no  clear change is observed in
the scale dependence of $\lambda_{\gamma p}(\ptm)$. Hence the 
evolution  towards the soft
region appears to be rather smooth.

%

\section{Conclusion}

A significant change is measured in the energy dependence of  
the partial cross section $\sigma_{\gamma p}(W,\ptm^2)$ with increasing transverse
momentum of the charged particles in the photon fragmentation region. 
This result cannot be described by a simple longitudinal phase space model.
It is, however, well reproduced by leading order QCD and can be 
qualitatively understood in the framework of Reggeon Field Theory.
A  similarity is observed
in the behaviour of the energy dependence
of the $\sigma_{\gamma p}(W,\ptm^2)$ and 
$\sigma^{tot}_{\gamma^*p}(W,Q^2)$. This
supports a common underlying picture for real and virtual photoproduction
as well as  a smooth and continuous transition from hard to soft processes.

\vspace{0.3cm}
\par\noindent
{\bf Acknowledgments}

\noindent
We are grateful to the HERA machine group whose
outstanding efforts made this experiment possible. We appreciate the immense
effort of the engineers and technicians who constructed and maintain
the H1 detector. We thank the funding agencies for financial support.
We acknowledge the support of the DESY technical staff. We wish to
thank the DESY directorate for the support and hospitality extended to
the non-DESY members of the collaboration.
We thank A. Kaidalov and E. Levin for stimulating discussions.

\end{document}